\documentclass[12pt]{article}
\usepackage{amsmath}
\usepackage{graphicx}
\usepackage{times}
\usepackage[T1]{fontenc}
\usepackage[utf8]{inputenc}
\usepackage[margin=1cm]{caption}

\usepackage[square,numbers,sort]{natbib}

\author
{Artem Litvinenko$^{1\ast}$, Roman Khymyn$^{1}$, Roman Ovcharov$^{1}$
and Johan \AA kerman$^{1,2,3\ast}$
\\
\normalsize{$^{1}$Department of Physics, University of Gothenburg, Fysikgränd 3, 412 96 Gothenburg, Sweden}\\
\normalsize{$^{2}$Center for Science and Innovation in Spintronics, Tohoku University, 2-1-1 Katahira,}\\
\normalsize{Aoba-ku, Sendai, 980-8577, Japan}\\
\normalsize{$^{3}$Research Institute of Electrical Communication, Tohoku University, 2-1-1 Katahira,}\\
\normalsize{Aoba-ku, Sendai 980-8577 Japan}\\
\\
\normalsize{$^\ast$To whom correspondence should be addressed; E-mails:} \\
\normalsize{artem.litvinenko@physics.gu.se, johan.akerman@physics.gu.se}
}

\date{}

\begin{document} 

\parindent 0cm
\parskip 12pt

\title{\LARGE\bfseries{A 50--spin surface acoustic wave Ising machine}} 

\maketitle

\abstract{Time-multiplexed Spinwave Ising Machines (SWIMs) have unveiled a route towards miniaturized, low-cost, and low-power solvers of combinatorial optimization problems. While the number of supported spins is limited by the nonlinearity of the spinwave dispersion, other collective excitations, such as surface acoustic waves (SAWs), offer a linear dispersion. Here, we demonstrate an all-to-all, fully FPGA reprogrammable, 50-spin surface acoustic wave-based Ising machine (SAWIM), using a 50-mm-long Lithium Niobate SAW delay line, off-the-shelf microwave components, and a low-cost FPGA. The SAWIM can solve any 50-spin MAX-CUT problem, with arbitrary coupling matrices, in less than 340 $\mu$s consuming only 0.62 mJ, corresponding to close to 3000 solutions per second and a figure of merit of 1610 solutions/W/s. We compare the SAWIM computational results with those of a 100-spin optical Coherent Ising machine and find a higher probability of solution. Moreover, we demonstrate that there is an optimum overall coupling strength between spins at which the probability of the exact solution reaches 100\%. The SAWIM illustrates the general merits of solid state wave-based time-multiplexed Ising machines in the microwave domain as versatile platforms for commercially feasible high-performance solvers of combinatorial optimization problems.}

\section*{Introduction}
Rapid and energy-efficient solvers of combinatorial optimization problems are essential for numerous application within finance \cite{Ibarra1975Knapsack}, circuit design \cite{Barahona1988CO}, drug discovery \cite{Earl2005parallelTempering}, operations \cite{Cerny1985ThermodynamicalTSP}, and scheduling \cite{burke2004state}. Many combinatorial problems are nondeterministic polynomial time (NP)–hard, or NP-complete, problems \cite{Barahona_1982} with a time-to-solution parameter that scales exponentially with problem size ($N$) when solved with classical von Neumann architectures. Although approximations and heuristic methods \cite{du1998handbook} allow for significant acceleration, the exponential scaling represents a monumental challenge from both hardware and algorithm perspectives. Fortunately, solving these problems can be effectively accelerated by mapping them onto analog Ising machines whose time-to-solution scales as $\sqrt{N}$ \cite{mohseni2022IMreview}.

Analog Ising machines (IM) have experienced dramatic progress in the recent decade, with a large number of successful novel implementations \cite{mohseni2022IMreview} that can be separated into two main architectures: \emph{i}) physical arrays of Ising spins \cite{albertsson2021ultrafastSHNOIM,houshang2022SHNOIM,Sutton2017SpintronicStochIM,Cai2020,johnson2011quantumIM,dutta2021ising,TWang2019,moy20221,si2023energy,aadit2022massively,grimaldi2023spintronicIMs}, and \emph{ii}) time-multiplexed Ising machines \cite{Honjo2021SciAdv100kCIM,Inagaki2016Sci2000CIM,McMahon2016Sci100CIM,yamamoto2017CIMquantumfeedback,Kako2020CIMwithErrFB,takata16bitCIMdelaylines2016,Haribara2016CIMPerformanceEvalDelayLines,marandi2014networkCIM4delaylines,cen2022large,bohm2019poor} based on propagating phase-binarized pulses. 
While IMs based on the physical array architecture provide the fastest time-to-solution parameter in simulations \cite{albertsson2021ultrafastSHNOIM,houshang2022SHNOIM,johnson2011quantumIM}, their computational capacity is strongly hampered by their limited connectivity due to interconnection problems. The connectivity bottle-neck was removed by the optical Coherent Ising machine that pioneered a time-multiplexed architecture where Ising spins are represented by propagating light pulses. Time-multiplexing allows for all-to-all connections between spins via external analog or digital schemes and allowed CIMs to successfully demonstrate continuous growth in the number of supported spins from 4 to a record 100,000 spins \cite{marandi2014networkCIM4delaylines,McMahon2016Sci100CIM,Inagaki2016Sci2000CIM,Honjo2021SciAdv100kCIM}. Nevertheless, despite this impressive scalability, CIMs remain limited to laboratory demonstrations due to their large physical size, high power consumption, poor temperature stability, and cost. 

These issues were recently addressed in the design of a Spinwave Ising Machine (SWIM) \cite{litvinenko2023spinwave,Gonzalez2023GlobalZTSWIM} where the Ising spins were represented by spinwave radio-pulses propagating at orders of magnitude slower speeds of km/s. Their reduced speed and low central frequency allowed for a dramatic miniaturization to mm dimensions and a much improved frequency stability. Here, we further push the concept of IMs based on solid-state collective excitations by demonstrating a time-multiplexed surface acoustic wave-based Ising Machine (SAWIM). The main limitation of the number of spins in SWIM was the nonlinearity of the spinwave dispersion \cite{serga2010YIGMagnonics} that leads to a non-constant delay time within the spinwave generation bandwidth, and significant pulse broadening towards the end of the delay line. In contrast, the SAW dispersion is intrinsically linear over a wide range of frequencies and amplitudes, which removes any significant broadening of the propagating SAW pulses. Altogether, the use of SAWs opens a path for miniature, low-power, and low-cost time-multiplexed Ising machines with all-to-all connectivity solving large-scale arbitrary combinatorial problems.

\section*{Surface Acoustic Waves}\label{sec2}
Surface acoustic waves (SAWs), first described by Lord Rayleigh \cite{rayleigh1885waves}, are a family of well-studied mechanical waves \cite{biryukov1995surface} propagating along the surface of a solid material and confined to a depth of about one wavelength. SAW devices typically employ piezoelectric materials, such as lithium niobate (LiNbO$_3$), as this allows for straightforward electrical transduction. The propagation of SAWs along a delay line can be described using a simple wave equation:

\begin{equation}
    \frac{\partial^2 u(x,t)}{\partial t^2} = v^2 \frac{\partial^2 u(x,t)}{\partial x^2} + S(x,t)
\end{equation}
\\
where $v$ is the phase velocity of the surface wave, $u(x,t)$ the displacement, and $S(x,t)$ a source term representing \emph{e.g.}~an interdigital transducer (IDT) \cite{malocha2004evolution,meltaus2005low,iriarte2012fabrication}. The very simple and linear SAW dispersion, $v = 2\pi f / \lambda$, where $f$ is the frequency and $\lambda$ the wavelength, ensures that microwave SAW pulses propagate without broadening. Their minimum duration is hence directly defined by the SAW bandwidth.

\section*{The Surface Acoustic Wave Ising Machine}\label{sec3}
\begin{figure*}[ht!]
\centering
\includegraphics[width=16cm]{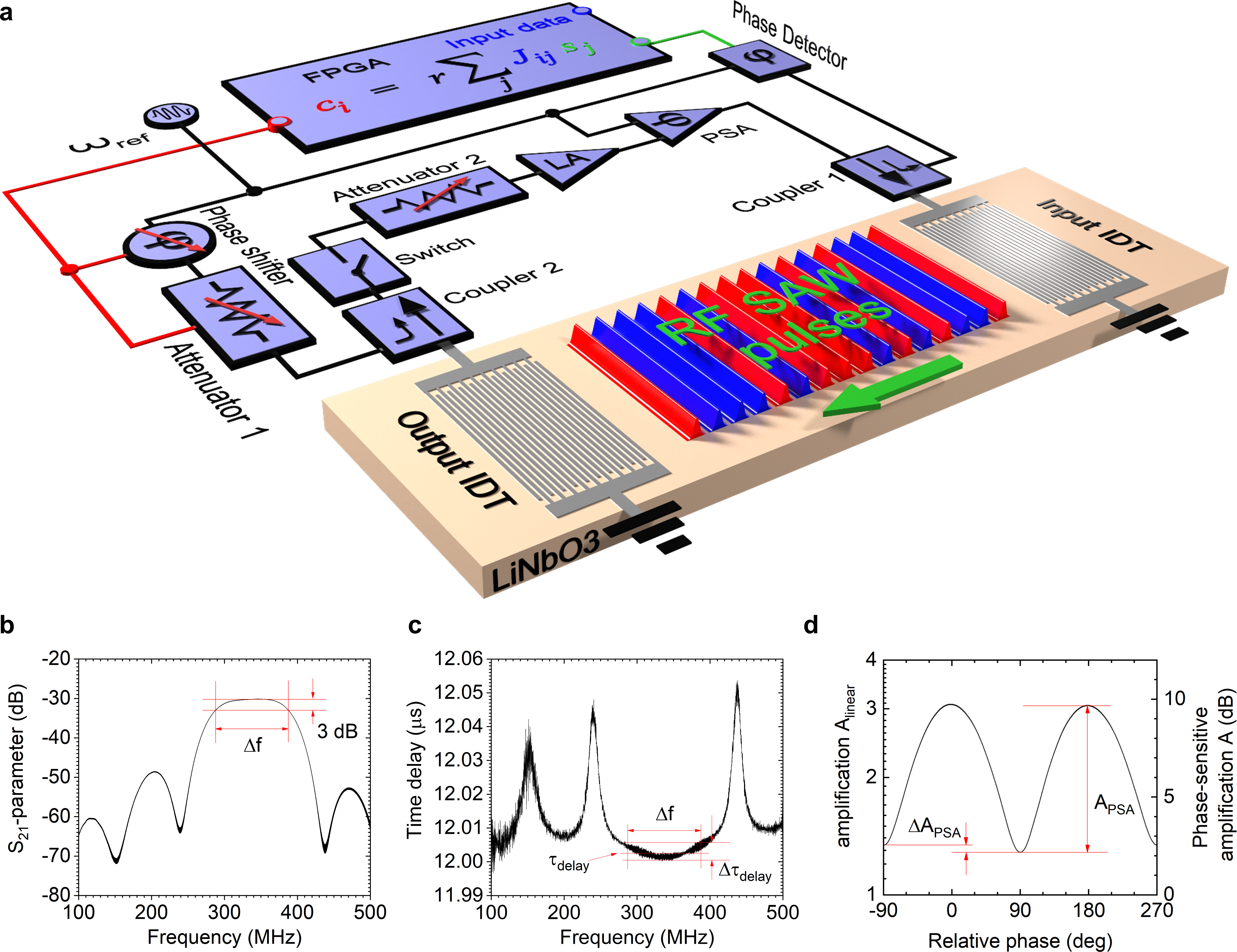}
\caption{\textbf{The surface acoustic wave Ising machine (SAWIM).}
(\textbf{a}) SAWIM schematic: LiNbO$_{3}$ is a substrate of the SAW delay line. Coupler 1 is used to deflect -15 dB of the propagating RF pulses to a Phase Detector. PSA is a parametric phase sensitive amplifier. LA is a microwave linear amplifier. Attenuator 2 is used to control the overamplification in the loop. Switch is forming 58 RF propagating pulses within 12$\mu s$ round trip time in the loop and has a control square signal with a frequency of 4.829 MHz and 50\% duty cycle. $\omega_{ref}$ is a reference signal of 320.118 MHz that is used by PSA, phase detector and phase shifter. FPGA is a digital measurement and feedback block that performs matrix multiplication between vector of spin values $s_{j}$ and coupling matrix $J_{ij}$ and controls coupling pulses with $c_{i}$ vector. Phase shifter has two states 0 and 180 degrees and controls the phase shift of the coupling pulses according to the sign of $c_{i}$ vector. Attenuator 1 controls the amplitude of the coupling pulses according to the magnitude of $c_{i}$ vector. Coupler 2 injects the coupling pulses to the propagating RF pulses with -15 dB of attenuation.
(\textbf{b}) Measurement of $S_{21}$ of the LiNbO$_{3}$ delay line. $\Delta f=$ 100 MHz is the acoustic wave spectral bandwidth measured at --3 dB.  
(\textbf{c}) The frequency-dependent delay time of the SAW delay line. $\tau_{delay}=$ 12$\mu s$ s is the average delay time over $\Delta f$ with a total deviation of $\Delta\tau_{delay}=$ 45 ns.
(\textbf{d}) Amplification of the parametric phase-sensitive amplifier as a function of the relative phase between the propagating RF pulses and the reference signal $\omega_{ref}=$ 6.25 GHz, with a phase sensitivity of $\mathit{A_{PSA}}=$ 6.1 dB.} 
\label{FigSAWIM} 
\end{figure*}

Figure~\ref{FigSAWIM}a shows a schematic of the SAWIM. The core is a SAW delay line with a 12~$\mu$s delay time, a total deviation of $\Delta\tau_{delay}=$ 45~ns, and a 3-dB bandwidth of 100~MHz. Altogether, this allows for excitation and detection of microwave pulses of 45~ns duration without dispersion-related distortions. However, the resonance mechanism of phase-sensitive amplification requires at least 10--20 periods 
per pulse, which imposes a stronger limitation of 30--60~ns on the minimum pulse duration. Thus, the maximum number of artificial Ising spins supported by the main ring circuit is 100--200, assuming a 50\% duty cycle. To simplify the measurement and the feedback circuit, we use a radio-frequency to intermediate frequency (RF/IF) Gain and Phase Detector AD8302, which has a phase output response time of 50-70~ns. For unambiguous readings from the phase detector, we allowed a pulse duration of 100 ns, which further reduces the number of spins to 58, where 50 spins are fully interconnected Ising spins, and the remaining eight propagating SAW pulses are used for an additional delay needed for the measurement and feedback system. 

The first step consists of establishing stable circulating SAW microwave pulses in the multi-physics ring circuit consisting of the delay line and the peripheral electronics. The SAW delay line has 30~dB of attenuation, which is easily compensated for by linear and phase-sensitive amplifiers. According to the Barkhausen stability criterion, the total phase accumulation in the ring circuit has to be an integer multiple of $2\pi$, which is ensured by a proper choice of the reference frequency $\omega_{ref}$. To realize bi-stable Ising spins, the propagating SAW pulses are phase-binarized using a phase-sensitive amplifier (PSA). We note that the phase sensitive amplification can potentially be implemented directly in the delay line using traveling wave excitation sources \cite{li2019traveling,li2019use,hagelauer2022microwave}, which could further miniaturize the system. 

Once stable circulating SAW pulses have been established, they have to be interconnected to realize an Ising machine with a particular Ising Hamiltonian. We have designed a measurement and feedback FPGA-based block (MFB) similar to the one used in a 100-spin CIM \cite{McMahon2016Sci100CIM}. The system includes a --15dB coupler (Coupler 1) to siphon off a --15dB portion of the circulating signal, which is then fed to an RF/IF Gain and Phase Detector AD8302, an 8-bit Analog-to-Digital converter AD9280, a Xilinx Spartan-6 XC6SLX9 FPGA with a matrix multiplication block, a phase shifter block, and an attenuator HMC472 (Attenuator 1). Finally, another --15-dB coupler (Coupler 2) injects the feedback coupling pulses ($c_i$) back into the main ring. 

The energy of an arbitrary coupled array of $N$ phase-binarized SAW pulses is proportional to the corresponding Ising Hamiltonian \cite{lucas2014ising}, 

\begin{equation}
    H(s_1,...,s_N) =  -\sum_{i<j}J_{ij}s_i s_j - \sum_{i=1} h_i s_i,
    \label{eq:ising-hamilt}
\end{equation}
\\
where $s_i=\pm 1$ corresponds to in-phase/out-of-phase states of the phase-binarized pulses, $J_{ij}$ is an arbitrary interconnection/coupling matrix, and $h_i$ is an individual local bias field. $J_{ij}$ controls the ratio between the amplitude of the coupling pulse from a single spins to the amplitude of the saturated propagating pulses. The same spin configuration $s_{i}$ represents a maximum number of cuts in MAX-CUT problem:

\begin{equation}
    C(\{s_i\}) = -\frac{1}{2} \sum_{1 \leq i < j \leq N} J_{ij} - \frac{1}{2} H(\{s_i\}).
\end{equation}

A specific MAX-CUT problem can be mapped onto the SAWIM by adjusting the pairwise coupling terms $J_{ij}$. Since the SAWIM is based on time-multiplexed oscillators with feedback, the value of the total coupling pulse combined from all the interconnected pulses to a single pulse should not exceed 20-30\% of the saturated amplitude of the circulating pulses as it would lead to over-coupling between RF SAW pulses and potentially chaotization of the oscillatory circuit. On the other hand, in the case when it is less than 2-3\%, and has a negative phase, the necessary switching of the targeting pulse will not happen due to a constant over-amplification of the circulating pulses in the ring circuit. This limitation can be avoided while solving MAX-CUT problems by keeping the smallest amplitude above the switching threshold of 2\% and chopping off the maximum amplitude below the top threshold of 30\%. We note that for more complex problems, such as the Travelling Salesman Problem \cite{hasegawa2021optimization} and the Channel Assignment Problem \cite{hirotake2020allocation}, where the couplings between Ising spins should have a higher resolution, special mapping protocols might need to be established.

The MFB calculates the necessary amplitudes and phases for the feedback coupling pulses ($c_i$) proportional to $J_{ij}$ and the instantaneous spin state via:

\begin{equation}
    c_{i} =  -r\sum_{j=1}^nJ_{ij} s_j
    \label{eq:feedback-ising}
\end{equation}
\\
where $r$ is the ratio between the amplitude of saturated pulses and the amplitude of the feedback from a single artificial Ising spin.

We note that implementing a time-multiplexed Ising machine with circulating SAW microwave pulses leads to several improvements as compared to both the CIM and the SWIM. Compared to the CIM, the central frequency of the SAWIM is $6 \times 10^{5}$ times lower, leading to improved thermal stability due to a much smaller phase accumulation in the ring. While it is sufficient to have a few periods of oscillation for a microwave pulse to define an Ising spin, the first CIM spins had about $1 \times 10^{6}$ periods, and the latest 100,000-spin CIM, with 30~ps light pulses, have about 6000 periods. Any further decrease in the number of periods for CIM spins is likely unfeasible because of the FPGA bottleneck. However, using solid-state delay lines based on spinwaves or acoustic waves with much lower oscillation frequency, is more efficient and leads to a much improved thermal stability by a factor of $1.6854 \times 10^{5}$. Consequently, while the 100,000-spin CIM requires an external cryogenic thermostat system, the SAWIM can operate at room temperature without any temperature control and phase-locking systems. Moreover, the total power consumption of the demonstrated SAWIM active blocks adds up to just 1.82~W and can potentially be further reduced to less than 0.5~W. Its low power consumption and benchtop design make SAWIM a commercially feasible contender for combinatorial solvers.

\section*{Results}\label{sec4}
\begin{figure*}[ht!]
\centering
\includegraphics[width=16cm]{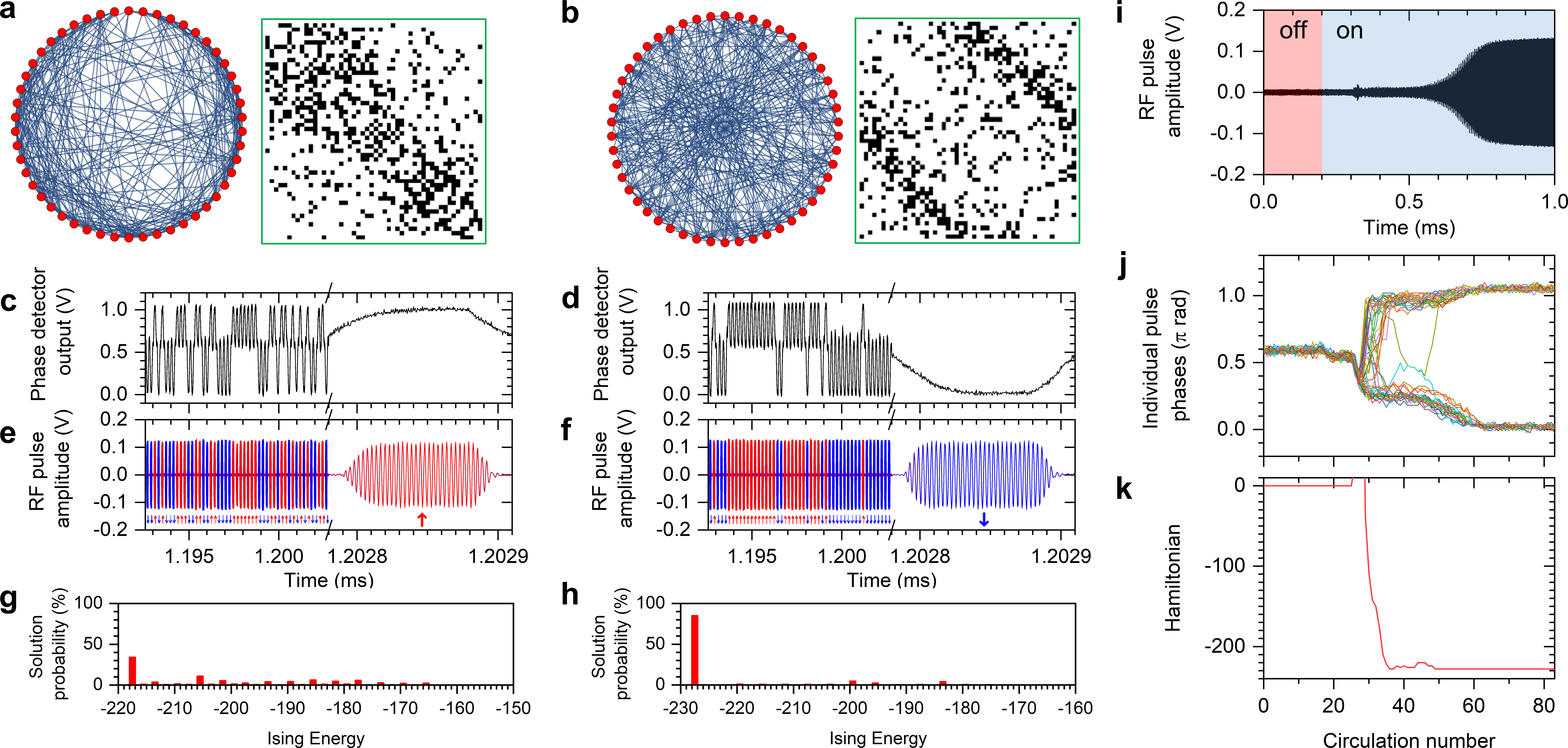}
\caption{\textbf{Experimental demonstration of the 50-spin SAWIM solving two different MAX-CUT problems.}
\textbf{a,b} Graph and chessboard representation of the solved MAX-CUT problems (\textbf{a}) Graph A (\textbf{b}) Graph B.
\textbf{c,d} Time traces of the phase detector output for Graph A and Graph B. 
\textbf{e,f} Time traces of the signal of RF pulse amplitude obtained from coupler 1. The time traces are colored according to the phase detector signal. The inset arrows demonstrate the corresponding up/down orientation of each quasi-spin. The data in \textbf{a,b,c,d,e,f} corresponds to an optimum solution of the MAX-CUT problems.
\textbf{e,f} The solution probabilities as a function of Ising Energy. 
\textbf{i} Time traces of the SAW pulse amplitudes in the beginning of the computation process. During the first 200 $\mu s$, the overall loop gain is below -20 dB and there are only thermal fluctuations. At 200 $\mu s$ the additional attenuation is switched off, the system becomes active as the overall amplification in the small signal regime becomes 2 dB.
\textbf{j} Temporal evolution of the individual discrete phase values of all 50 spins. 
\textbf{j} Temporal evolution of the Ising Energy of the Hamiltonian. The data in \textbf{i,j, k} correspond to the random matrix B. }
\label{FigMAXCUT} 
\end{figure*}

We have evaluated our Ising machine with different MAX-CUT Ising problems with $J_{ij} = {-1,0}$ and no Zeeman term, $h_{i} = 0$. In Fig.~\ref{FigMAXCUT}a, b, we show two random graphs, A and B, for two MAX-CUT problems with the corresponding chessboard representation of the $J_{ij}$ matrices, where black pixels correspond to $J_{ij} = -1$. Matrix A has $N$ = 50 vertices, $\lvert E \rvert = 301$ edges, and a density $d$ = 0.2457, whereas matrix B has 50 vertices, 262 edges, and a density of 0.2139. The time traces for the phase detector signal and the microwave pulse amplitude taken from the phase detector and are shown in Fig.~\ref{FigMAXCUT}c,e for matrix A and in Fig.~\ref{FigMAXCUT}d,f for matrix B. 
For consecutive computations, we turn the amplification in the SAWIM loop on and off (see Fig.~\ref{FigMAXCUT}i) by an additional digital attenuator (attenuator 2) with a square signal having a 10-ms period and a 95\% duty cycle to ensure that between computation cycles the SAW echo pulses completely decay (see Fig.~\ref{FigMAXCUT}j). 

We can see from Fig.~\ref{FigMAXCUT}j that most changes in the spin states happen already when the amplitude of the pulses is relatively small. Since the amplitude of the coupling pulses only depends on the spin configuration when the circulating spins have small amplitudes, the effective coupling is strong. Consequently, as the amplitude of the circulating pulses gradually increases, the effective coupling decreases, or, alternatively, the equivalent temperature of the system decreases, and, hence, the spin state freezes. Figure Fig.~\ref{FigMAXCUT}k shows the corresponding temporal evolution of the Hamiltonian. The solution probabilities for matrices A and B are shown in Fig.~\ref{FigMAXCUT}g,h. In both cases, the state with the highest probability corresponds to the optimum Hamiltonian, 34\% of H = --218 for matrix A and 85\% of H = --228 for matrix B. 

It takes 28 circulations, or around 340 $\mu s$, to achieve the absolute minimum of the Hamiltonian of --228 (see Fig.~\ref{FigMAXCUT}k) for the system with the corresponding coupling matrix B. Taking into account the SAWIM power consumption of 1.82~W, the required energy to find a solution for a 50-spin MAX-CUT problems is hence just 0.62~mJ. 

\begin{figure*}[ht!]
\centering
\includegraphics[width=16cm]{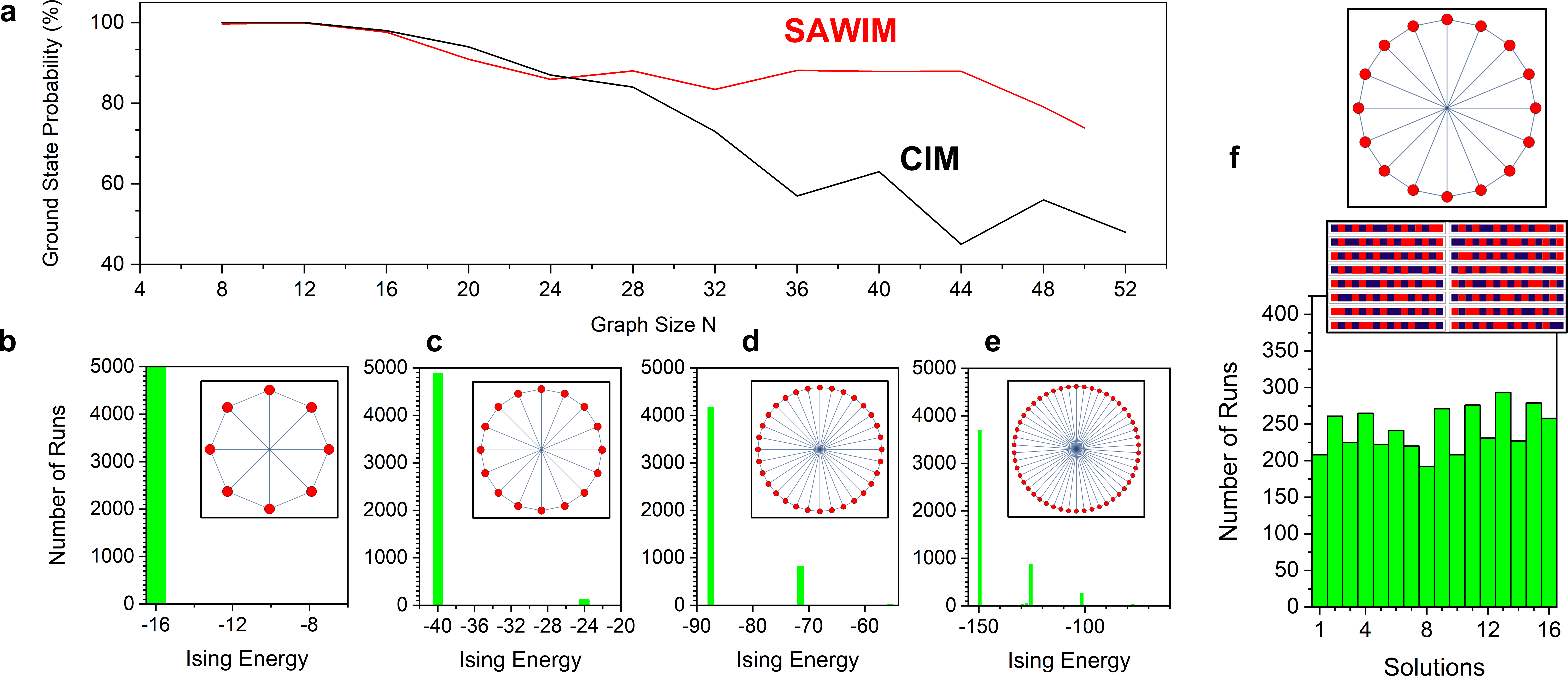}
\caption{\textbf{Statistics for the MAX-CUT problem with Möbius ladder graphs of various sizes.}
\textbf{a} Ground state probability for Möbius ladder graphs of size $N$ obtained from 5000 consecutive runs. 
\textbf{b,c,d,e} Distributions of the solutions by Ising Energy of the system Hamiltonian for 8-spin (\textbf{b}), 16-spin (\textbf{c}), 32-spin (\textbf{d}) and 50-spin (\textbf{e}) Möbius ladder graphs. The insets show the corresponding graph configuration.  
\textbf{f} Distribution of the solutions by the states with the Ising Energy of --40 for a 16-spin (\textbf{e}) Möbius ladder graph. The inset shows the corresponding graph configuration and all 16 possible spin states.}
\label{FigMobius} 
\end{figure*}

To benchmark our system against the 100-spin CIM, we also characterize the probability to find the ground state for Möbius ladder graphs of different sizes (8--50). We use 5000 system runs of 10 ms each to accumulate the statistics. Similarly to the CIM, the ground state probability starts to decay from 100\% at a graph size of 12. The distributions of the solutions by the Hamiltonian states for 8-spin, 16-spin, 32-spin, and 50-spin MAX-CUT problems are shown in Fig.~\ref{FigMobius}b,c,d,e. Compared to the 100-spin CIM, the ground state probabilities for graph sizes from 28 to 50 are significantly higher, which we attribute to a better phase stability in the system as well as differences in the strength of the coupling between artificial Ising spins. 

An important property of the Ising machine is the identity of the artificial spins, their coupling strengths, and the absence of any preferred direction for any spin, potentially arising from parasitic global or local Zeeman terms. To confirm that the SAWIM has no such bias, we performed 5000 runs with a 16-spin Möbius ladder graph coupling scheme. Due to the symmetry of the Möbius ladder graph, there are 16 states with the same ground state energy --40 (see the insets of Fig.~\ref{FigMobius}c,f). It can be seen from Fig.~\ref{FigMobius}f that the distribution of all the solutions by 16 possible states is mostly uniform. It confirms that the SAW Triple Transit Echo (TTE) level is low, and that there are no unintended couplings between the nearby circulating SAW pulses.

\subsection*{Optimal Coupling}\label{sec5}
As suggested in \cite{bashar2023stability}, there exists an optical coupling strength for oscillator based IMs: if the coupling is too weak, the system does not reach a stable solution; if the coupling is too strong, non-optimal, local minima, might hinder the IM from reaching its ground state. 
Since the SAWIM is constructed with electronic microwave amplifiers, which are intrinsically nonlinear, there is an optimum level of coupling between spins when the probability of the optimum solution is maximal. To verify the applicability of this theoretical prediction to time-multiplexed Ising machines, and particularly to our SAWIM, we introduced a numerical toy model, where each time-multiplexed Ising spin can be considered as a separate weakly nonlinear oscillator with a delay in a feedback loop. The evolution of the complex amplitude $a_j$ of the $j^{\text{th}}$ spin, can be described by the following equation \cite{tiberkevich2014sensitivity}, which have been demonstrated to be applicable to time-multiplexed IMs \cite{litvinenko2023spinwave}:
\begin{equation}
\frac{da_j}{dt}+(i \omega_0+\Gamma_0)a_j-K\left(1-\beta_r\delta p_{j,\tau}\right) \exp\left(-i\beta_i\delta p_{j,\tau}\right)a_{j}(t-\tau)=\xi(t)
\label{eq:delay}
\end{equation}
where $\delta p_{j,\tau}=(p_j(t-\tau)-p_s)/p_s$, $\tau=\tau_{delay}$ is a delay time along the loop, $\Gamma_0$ describes losses, $\omega_0$ is the resonance frequency, $K$ the gain, $\beta_r$ the positive coefficient of the gain compression, $\beta_i$ the nonlinear frequency shift, and $p(t)=\lvert a(t) \rvert^2$ the power of oscillations with an operating point at $p_s$. The right hand side of Eq. (\ref{eq:delay}), $\xi(t)=K_e\exp^{i\omega_e t}a_j^*+\kappa \sum_{i} J_{ij}a_{i}/\lvert a_{i} \rvert$, contains signals from the outer loops, \emph{i.e.}~the synchronization signal at the double frequency $\omega_e \simeq 2\omega_0$ and couplings from other spins, where $\kappa$ is a coupling coefficient. For the simplicity of calculations, we chose the following parameters: $\omega_0/2\pi=1.0$, $\omega_e/2\pi=2.002$, $\tau=10.0$, $\Gamma_0/2\pi=4.0 \times 10^{-2}$, $\beta_r=2 \times 10^{-2}$,  $\beta_i=1.0 \times 10^{-4}$, $K/2\pi=4.0 \times 10^{-2}$, $K_e/2\pi=6.0 \times 10^{-3}$.

\begin{figure*}[ht!]
\centering
\includegraphics[width=9cm]{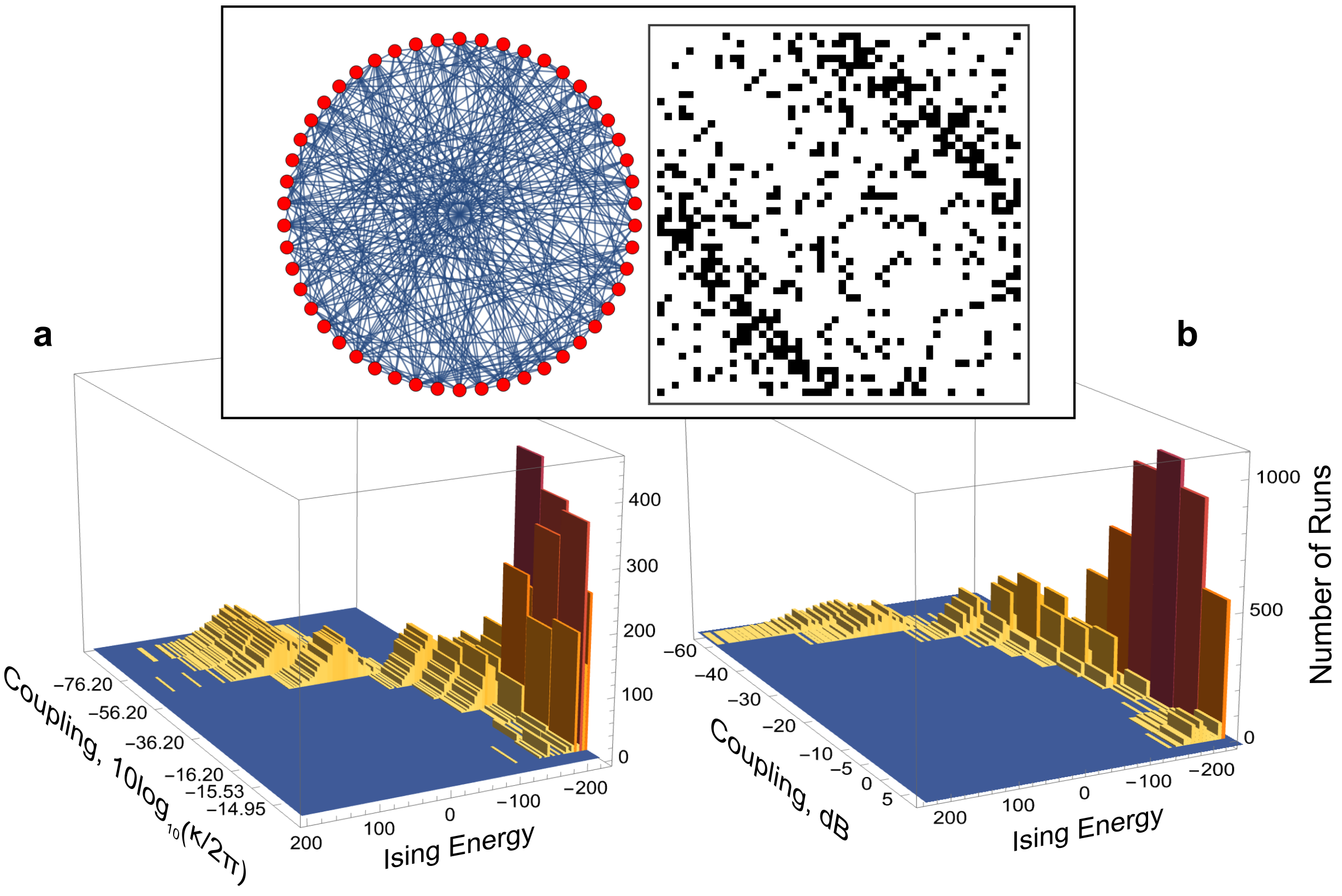}
\caption{\textbf{Characterization of the optimal strength of the coupling for the matrix B}. 
Distribution of the solutions at the states with different Ising energy: a) for the numerical model, b) produced by the hardware SAWIM.  The additional attenuation of coupling strength is indicated with the labels on the left side. The inset on top shows the graph and chessboard representation of $J_{ij}$ coefficients of the matrix B. The results are presented for 1000 runs per each coupling attenuation.}
\label{FigOptimalCoupling} 
\end{figure*}

The results of the calculations are shown in Fig.~\ref{FigOptimalCoupling}a. As expected, low values of the coupling strength makes the probability to reach the ground state negligibly small, while for strong coupling the IM exhibits an elevated risk of getting stuck in local minima, which also reduces the probability of finding the optimal solution. Thus, our calculations emphasize the need to find optimal coupling strengths, in agreement with \cite{bashar2023stability}.

To corroborate these results with experiments, we performed a series of SAWIM runs with different global attenuation of the coupling strengths. Fig.~4b shows how the distribution of the Ising energy of the solutions of the MAX-CUT problem defined by matrix B changes with coupling strength. The top distribution corresponds to zero coupling, resulting in random spin configurations with a wide Gaussian distribution of the Ising energies. As the coupling strength increases, the distribution width narrows sharply and the mean rapidly moves to lower energies. At --5 dB of attenuation, 100\% of the solutions are optimal at --228. However, when the coupling strength is instead amplified by 5 dB, the distribution of the solutions again spreads to higher Ising energies, but now with a more scattered, non-Gaussian, distribution, reflecting how the system gets stuck in a few local minima. Our experiment hence confirms that the time-multiplexed SAWIM has an optimum global coupling strength at which the stability of the ground state is maximized, reaching a 100\% probability.

\section*{Scalability}
We demonstrated our 50-spin SAWIM using a 50 mm long SAW delay line with a 12-$\mu$s delay, which could theoretically accommodate up to 200 spins. To further scale up the number of spins, one could explore different directions, such as increasing the delay time, increasing the central frequency ($f$), and reducing the SAW pulse width. As the acoustic losses exhibit a $f^2$ dependence, lower central frequencies can result in 1~ms and even 10~ms delay time \cite{Fortunko1974,Coldren1976}; operating at 100MHz with a delay time of 1--10~ms could hence make the number of spins approach $10^4-10^5$. On the other hand, low frequencies and long delay times lead to a corresponding increase in the time-to-solution and, therefore, the use of SAW delay lines operating at GHz range is equally interesting. SAWs can be excited and propagate efficiently with acceptable losses at frequencies up to 4-6~GHz \cite{makkonen2003surface,plessky2014saw}, which would allow for 2-3~ns SAW pulses, and with $\mu$s-delay times could also reach thousands of equivalent Ising spins.

Once the maximum number of spins per delay line has been reached, one may explore the use of multiple delay lines, either in series or as parallel loops with individual linear and phase-sensitive amplifiers. In the case of series interconnection, the use of intermediate electronic or integrated SAW PSAs \cite{li2019traveling,li2019use,hagelauer2022microwave} would keep the SNR of the propagating SAW RF pulses similar to the present study, and the temperature stability would also remain the same as the intermediate PSAs correct the phase of the circulating RF pulses. The advantage of series interconnection is the possibility of using a single MFB system. However, the time-to-solution parameter will rise in proportion to the circulation time. In contrast to the series interconnection scheme, the system of parallel loops would keep the time-to-solution mostly proportional to the circulation time of a single loop, at the cost of multiple MFBs. Using a combination of these approaches, the SAWIM hence has the potential for scalability up to $10^4-10^6$ spins, while retaining its advantageous high-level temperature stability and high dynamic range.

Finally, the SAWIM has the same architecture as the latest CIMs and can operate under the minimum gain principle \cite{yamamoto2017CIMquantumfeedback}. The amplification in the loop can be gradually increased with voltage-controlled analog or digital attenuators giving enough time for the system to start up at the lowest energy state corresponding to the Ising Hamiltonian's minimum state.

\section*{Performance comparison with other Ising machines}
\begin{figure*}[ht!]
\centering
\includegraphics[width=16cm]{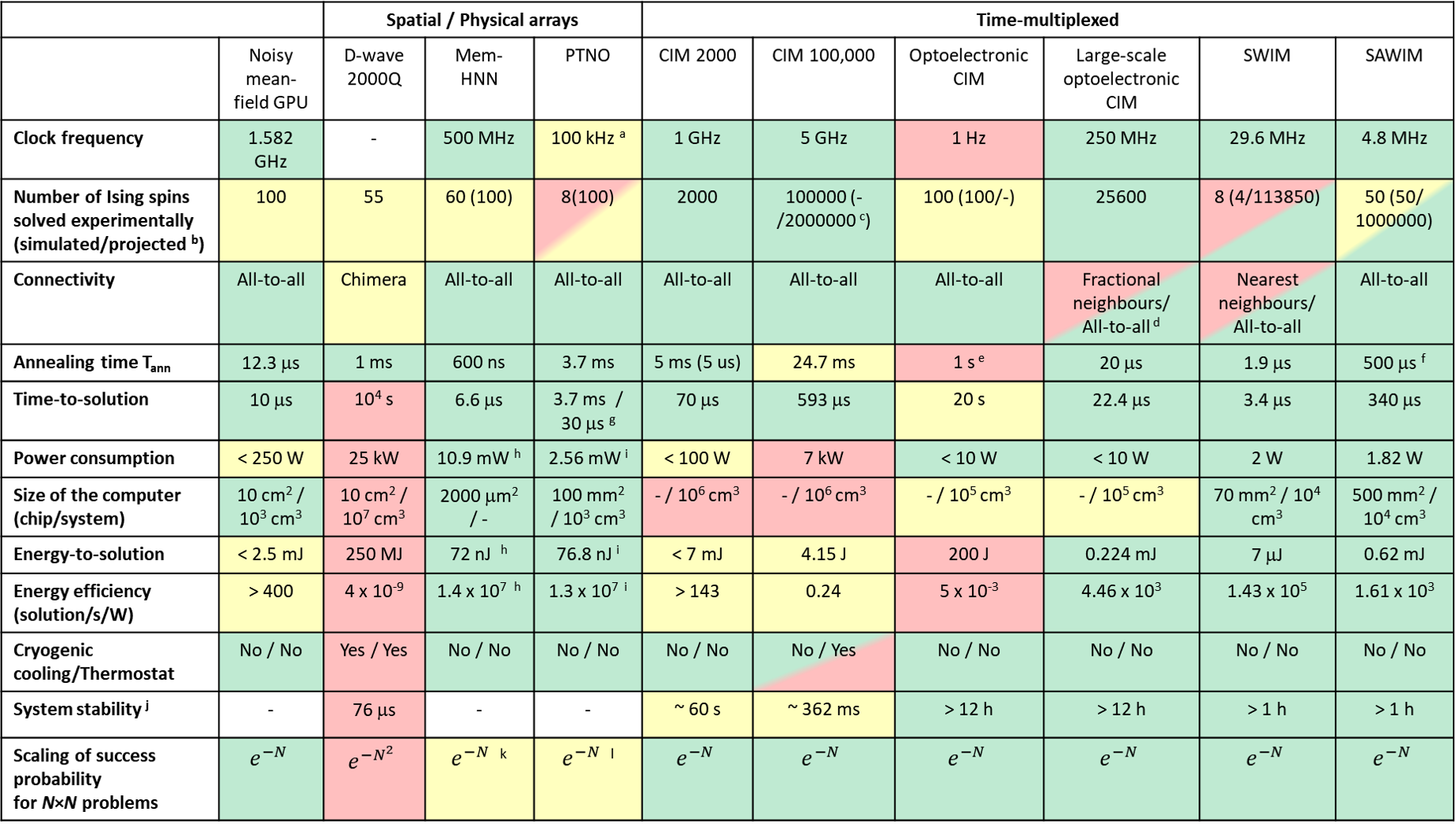}
\caption{\textbf{Comparison of the SAWIM and the current state-of-the-art annealing accelerators and Ising machines}. 
$^{a}$ Oscillation frequency of phase-transition oscillators.
$^{b}$ Projected number of artificial spins according to the potential in scalability 
$^{c}$ In ref. \cite{haribara2016computational}.
$^{d}$ Several delay lines with different delay bits allows to map onto optoelectronic CIM a limited number of MAX-CUT problems. However, a large-scale OE-CIM allows all-to-all connection with an FPGA system similar to 100000-spin CIM, OE-CIM and SAWIM.
$^{e}$ Mostly limited by the data acquisition interface and potentially can be reduce by 3-5 orders of magnitude.
$^{f}$ 500 $\mu s$ is the time to reach saturation in propagating SAW RF pulses. Since the amplitude of the coupling pulses only depends on $J_{ij}$ and the phases of the propagating SAW RF pulses $s_{i}$, the settling time of 500 $\mu s$ is identical to the annealing time.
$^{g}$ 30 $\mu s$ is simulation while in experiment only 3.7 ms was achieved.
$^{h}$ Power consumption and energy efficient were simulated.
$^{i}$ Power consumption and energy efficient were calculated for simulated PTNO IM with 100 spins without taking into account the power consumption of the control PC.
$^{j}$ Coherence times for the spins in the steady state
$^{k,l}$ Despite $e^{-N}$ scaling of success probability for small problems, the interconnection issues at large number of equivalents Ising spins may require the use of Chimera or sparse interconnection schemes that affects scaling of success probability.
}
\label{FigBencmarking} 
\end{figure*}

In order to evaluate the SAWIM position within a wide class of physical Ising machines, we benchmarked our SAWIM with other state-of-the-art implementations of Ising machines (Fig.~5). We have picked relevant metrics that affects the throughput of the system (connectivity, time-to-solution), the energy efficiency (system power consumption and energy required to reach solution), system stability (the requirement of cryogenic systems and thermostats, coherence times for spins), system size and potential for scalability (connectivity, scalability of the success probability and projected number of spins). For a wide-scale comparison, we included in the benchmarking the state-of-the-art commercially available solution D-Wave 2000Q quantum annealer, containing 2,048 qubits \cite{johnson2011quantumIM,hamerly2019experimental}, and a noisy mean-field annealing algorithm running on an NVIDIA GeForce GTX 1080 Ti GPU \cite{king2018emulating}. For highlighting solutions based on different physical principles but having a high potential for CMOS integration we considered a discrete-time memristor-based hybrid analogue–digital accelerator implementing a Hopfield neural network (mem-HNN) \cite{Cai2020} and an Ising Hamiltonian solver based on coupled stochastic phase-transition nano-oscillators \cite{dutta2021ising}. The last five IMs are from the same class of time-multiplexed Ising machines: a 2000-node Coherent Ising Machine \cite{Inagaki2016Sci2000CIM}, a state-of-the-art 100,000 spin Coherent Ising Machine \cite{Honjo2021SciAdv100kCIM}, and alternative simplified optoelectronic CIM \cite{bohm2019poor}, and large-scale multi-wavelength optoelectronic CIM \cite{cen2022large}, SWIM \cite{litvinenko2023spinwave} and, finally, SAWIM. The SAWIM has a comparable or even better time-to-solution parameter relative to the first four candidates with the potential of sufficient increase in the number of spins inherent for time-multiplexed systems.

The absence of physical delay systems in seemingly simplified optoelectronic CIMs \cite{bohm2019poor} in fact requires a more complex FPGA control system \cite{bohm2019poor} than in classical CIMs \cite{Inagaki2016Sci2000CIM,Honjo2021SciAdv100kCIM} and which has to store the spins amplitude and phases with high precision that significantly limits the maximum number of Ising spins. In the case of large-scale optoelectronic CIM \cite{cen2022large}, the absence of a physical delay system is compensated by the use of constant delay-time delay lines that dramatically limit the variety of possible encoded problems, essentially limiting the large-scale optoelectronic CIM to laboratory demonstrations without the possibility of solving arbitrary problems.

As compared with classical CIMs \cite{Inagaki2016Sci2000CIM,Honjo2021SciAdv100kCIM}, the use of a solid-state SAW delay line operating at a central frequency of 320 MHz results in a high thermal stability that is by a factor of $1.6854 \times 10^{5}$ better than in CIMs \cite{Inagaki2016Sci2000CIM,Honjo2021SciAdv100kCIM}. In contrasts to CIM, this allowed for operation without a phase locked loop (PLL). However, for even higher stability in larger and commercial SAWIM systems, we anticipate using simple PLL circuits.

We note that our SAWIM has a power consumption of just 1.82 W and an energy efficiency of $1.61 \times 10^3$ solutions per second per watt, which is comparable with optoelectronic CIMs and several orders of magnitude better than in other candidates such as D-Wave and CIM while still having all-to-all connectivity and the scalability potential of CIM. Moreover, with dedicated CMOS circuits, the power consumption of microwave blocks can be reduced to $\mu$W level. Finally, SAWIM is smaller than most systems, and has a good prospect for CMOS integration.

\section*{Conclusion}\label{sec12}
We have demonstrated the design of a time-multiplexed surface acoustic wave Ising machine (SAWIM). The SAWIM has a circulation time of 12 $\mu$s and a time-to-solution of about 340 $\mu$s for 50-spin MAX-CUT problems, consuming only 0.62 mJ to compute a single solution. We have benchmarked our SAWIM using different size Möbius ladder graphs and demonstrated its advantages as compared with a corresponding 100-spin CIM in the literature. We have also demonstrated the importance of an optimized coupling strength at which the ground state probability can reach 100\% for 50-spin MAX-CUT problems. The SAWIM consumes just 1.82~W of power and has significant potential for further power reduction. Our work presents a versatile, low-power, fast, and cheap microwave-based time-multiplexed Ising machine platform, with potential for dramatic miniaturization compared to optical solutions. The high SNR and dynamic range of coupling strength will allow for the implementation of high-resolution coupling matrices, enabling mapping of more complex problems, such as the travelling salesman problem, the nurse scheduling problem, and channel allocation problems. The demonstrated concept offers a clear path for scalability to thousands and potentially hundreds of thousands of spins. Finally, the operation at room temperature and the use of off-the-shelf low-power electronic components promises a high potential for commercialization. 

\section*{Methods}\label{sec13}
\subsection*{Sample}
The lateral dimension of the SAW waveguide is 50 x 10 mm$^2$. The microwave-to-acoustic wave transducers are silver IDTs. The substrate is Lithium Niobate. 

\subsection*{Electrical measurements}
The group delay presented in Fig.~\ref{FigSAWIM}c was obtained from phase-delay measurement of the S-parameters using a vector network analyzer. The S-parameters were measured across a wide range of frequencies starting from 100 MHz to 500 MHz with a resolution of 10 kHz and 40000 points. The group delay is calculated as the negative derivative of the phase shift with respect to the frequency. This calculation was performed using a built-in function of a vector network analyzer Rohde \& Schwarz ZNB 40. The time delay plot was smoothed with an FFT filter with 10-points integration.

The phase-sensitive amplification curve was extracted from the envelope of an amplified signal with constant amplitude and frequency detuned from the reference signal frequency by 10 MHz. The frequency detuning introduced a continuous phase drift with a period of 1 $\mu$s, which allowed a smooth phase-sensitive amplification curve to be extracted.

\section*{Author contributions}
A.L. conceived the concept and designed the circuit; A.L. performed the measurements and analyzed the data; R.K., R.O., performed theoretical calculations; J.Å. managed the project; all co-authors contributed to the manuscript, the discussion, and analysis of the results.

\section*{Acknowledgements}

This work was supported by a Knut and Alice Wallenberg Foundation WALP grant, KAW 253129326, a Horizon 2020 research and innovation programme ERC Advanced Grant No.~835068 "TOPSPIN", an ERC Proof of Concept Grant No.~101069424 "SPINTOP", and the Marie Skłodowska-Curie grant agreement No.~101111429 "SWIM".

\bibliography{references}

\end{document}